\documentstyle[12pt]{article}
\begin{document}

\hyphenation{mono-pole  mono-poles}
\title{The doublet of Dirac fermions in the field of the non-Abelia
monopole  and parity selection rules}
\date {}
\medskip

\author {V.M.Red'kov \\
     Institute of Physics, Belarus Academy of Sciences\\
F Skoryna  Avenue 68, Minsk 72, Republic ob Belarus\\
e-mail: redkov@dragon.bas-net.by}

\maketitle


\begin{abstract}

The paper concerns  a  problem  of  Dirac  fermion  doublet  in  the
external monopole potential arisen out of embedding the Abelian  monopole
solution in the non-Abelian scheme. In this particular case, the  Hamiltonian
is invariant under some symmetry  operations  consisting  of  an  Abelian
subgroup in the complex rotational group
$SO(3.C) : [\hat{H} , \hat{F}(A)]_{-} = 0 , \hat{F}(A) \in  SO(3.C)$.
This symmetry results in a certain  freedom  in  choosing  a
discrete operator entering the complete set
$\hat{H}$ , $\vec{j}^{2}$ , $j_{3}$,
$\hat{N}_{A} = [(-i) \exp  (-i A \vec \sigma \vec{n}_{\theta ,\phi })
\otimes (i \gamma ^{0}) \otimes  \hat{P} ] , \hat{K}.
$
The  same  complex
number  $A$   represents a parameter of the basis wave functions
$\Psi ^{A}_{\epsilon jm\delta \mu }(t,r,\theta ,\phi )$   constructed.
The  {\em generalized}  inversion-like  operator
$\hat{N}_{A}$  implies its own ($A$-dependent) definition for scalar and
pseudoscalar, and further affords some generalized $N_{A}$-parity
selection rules.

It is shown that all different sets of basis functions
$\Psi ^{A}_{\epsilon jm\delta \mu }(x)$ determine the same Hilbert space.
In particular,  the  functions $\Psi ^{A}_{\epsilon jm\delta \mu }(x)$
decompose into linear combinations of
$\Psi ^{A=0}_{\epsilon jm\delta \mu }(x)$. However, the bases considered
turn out to be nonorthogonal ones when $A^{*}\neq A$; the latter correlates
with the non-self-conjugacy property of the operator  $\hat{N}_{A}$
at  $A^{*}\neq  A$.

(This is a shortened version  of the paper).
\end{abstract}

\newpage
\subsection*{1. Introduction}

Together with geometrically topological way of classifying
monopoles, another approach to studying various monopole
configurations is possible, which concerns manifestations
of mo\-no\-poles playing the role of external potentials. Moreover,
from the physical standpoint the latter method can be thought of  as
a more visualizable one in comparison with less obvious and more
complicated topological language. So, the basic frame of the present
investigation is an analysis of particles in the external monopole
potentials. Here, both  Abelian  and  non-Abelian  cases will be
discussed (see also [1-8, 9-13]), although the non-Abelian case
is of primary interest to us.

Instead of the so-called monopole harmonics [4-8],  the  more
conventional  formalism of the Wigner's $D$-functions is used. In
contrast with the wide-spread  approach,  the  general  relativity
tetrad formalism of Tetrode-Weyl- Fock-Ivanenko  [14]  is  applied
to this problem. This enable us to reveal explicitly a connection
between the monopole topic and the early Pauli's investigation [15]
about the problem of allowable spherically symmetric wave functions
in quantum mechanics; his results bear on the Dirac's $eg$-quantization
condition [16] ($e$  and $g$  are respectively the electric and magnetic
charges). In addition, this Pauli's work is of interest to us because it,
among other things, used (heuristically) a special tetrad basis which can be
associated with the unitary isotopic gauge in the non-Abelian
monopole problem.

So, the major content  of the work reported in  this  article
is motivated by this Pauli's paper [15] and foregoing special
tetrad basis (it was introduced into the literature by Schr\"odinder
[17]; see also [18]). It should be noted that some generalized
analog of the Schr\"odinger's basis may be successfully used,
whenever in a linear problem there exists a spherical symmetry,
irrespective of the concrete embodiment of this symmetry.

\subsection*{2. The Dirac and Schwinger unitary gauges in the isotopic
        space and monopole potentials}

It is well-known that the usual Abelian monopole potential
generates a certain non-Abelian potential being a solution of the
Yang-Mills (Y-M) equations. First, such a~specific non-Abelian
solution was found out in [19]. A procedure itself of that
embedding the Abelian monopole 4-vector $A_{\mu }(x)$
in the non-Abelian scheme:
$  A_{\mu }(x) \rightarrow  A^{(a)}_{\mu }(x)  \equiv
   ( 0 , 0 , A^{(3)}_{\mu }$ = $ A_{\mu }(x) )$
ensures automatically that   $A^{(a)}_{\mu }(x)$    will satisfy
the free Y-M equations. Thus, it may be readily verified that the
vector
         $A_{\mu }(x) = (0, 0, 0, A_{\phi } = g\cos\theta )$
obeys the Maxwell general covariant equations in every curved
space-time  with  the  spherical symmetry:
$
dS^{2}=[e^{2\nu }(dt)^{2} - e^{2\mu }(dr)^{2} -
r^{2} ((d\theta )^{2} +\sin^{2}\theta  (d\phi )^{2}) ] $ ;
$
A_{\phi } =  g\cos\theta  \rightarrow
F_{\theta \phi } = - g \sin\theta;
$
here we get essentially a single equation
${{\partial} \over {\partial \theta}} [ \sin\theta
( - g \sin \theta ) / \sin^{2} \theta] = 0$.
In turn, the non-Abelian tensor   $F^{(a)}_{\mu \nu }(x)$    defined  by
$
F^{(a)}_{\mu \nu }(x) =  \nabla _{\mu } A^{(a)}_{\nu } -
\nabla _{\nu } A^{(a)}_{\mu } + e \epsilon _{abc} A^{(b)}_{\mu }A^{(c)}_{\nu }
$
and associated with the $A^{(a)}_{\mu }$ above  has a  very  simple  isotopic
structure:  $ F^{(3)}_{\theta \phi } = - g \sin\theta $   and  all
other   $F^{(a)}_{\nu \mu }$    are  equal  to
zero. So, this substitution
$
F^{(a)}_{\nu \mu } =
( 0 , 0, F^{(3)}_{\theta \phi } = - g \sin \theta)
$
leads the Y-M equations to the single equation of  the  Abelian case.
Strictly speaking, we cannot state that $A^{(a)}_{\mu }(x)$ obeys  a
certain set of really nonlinear equations (it satisfies linear
rather than nonlinear equations). Thus, this monopole potential
may be interpreted as a trivially non-Abelian solution  of Y-M
equations (this holds in every space-time with the spherical symmetry,
but the case of ordinary flat space will be of primary interest to us).

Supposing that such a sub-potential is presented in the
well-known monopole solutions of t'Hooft-Polyakov (and Julia-Zee) [20-22] :
$$
 \Phi ^{(a)}(x) = x^{a} \Phi (r) , \qquad
 W^{(a)}_{0}(x) = x^{a} F(r)     , \qquad
 W^{(a)}_{i}(x) = \epsilon_ {iab} x^{b} K(r)
\eqno(1)
$$

\noindent we can try to establish explicitly that constituent structure.
The use of the spherical coordinates and special gauge transformation
allows us to separate the trivial and
non-trivial parts of the potentials (1) into  different isotopic components
$ W^{1}_{\mu} , W^{2}_{\mu} , W^{3}_{\mu} $:
$$
W^{S.(a)}_{\theta} = \left( \begin{array}{c}
                     0\\ (r^{2}K + 1/e)\\  0
                     \end{array}\right); \qquad
W^{S.(a)}_{\phi}  =  \left( \begin{array}{c}
                     -(r^{2}K + 1/e) \sin \theta   \\   0
                     \\           {1\over e} \cos\theta
                     \end{array}\right)
\eqno(2)
$$

\noindent the symbol S. stands for the Schwinger unitary gauge in isotopic
space. To the above-mentioned special monopole field there
corresponds the  $K(r) = - 1/er^{2}$, so that the relations  from (2)
turn out to be very simple and related to the Abelian potential
embedded into the non-Abelian scheme.

\subsection*{3. Diagonalized operators and separation of variables}

In Section 3 we enter on analyzing the  isotopic doublet of
Dirac fermions in the external t'Hooft-Polyakov monopole field.
We are going to reexamine this problem all over again, using
the general relativity formalism [14]. Given the specified  tetrad
basis (Schr\"odinder basis of  spherical tetrad [17,18]) and the
unitary Schwinger frame in the local isotopic space (see (2)),
the matter equation takes the form
$$
\left [ \; \gamma ^{0} \;( i \partial _{t} + e r F t_{3} ) \; + \;
i \gamma ^{3} \; ( \partial _{r} + {1\over r} ) \; + \;
{1\over r} \Sigma ^{S.}_{\theta ,\phi } \; +    \right.
$$
$$
\left. +\; {{e r^{2}K + 1}\over r}\; (\gamma ^{1} \otimes  t^{2} -
\gamma ^{2} \otimes  t^{1}) \; - \;
( m + \kappa   r \Phi  t_{3} )\;  \right ]  \;
\Psi ^{S.} = 0 \; ;
\eqno(3)
$$
$$
\Sigma ^{S.}_{\theta ,\phi } = \left ( i \gamma ^{1} \partial _{\theta }\; + \;
{{i\partial _{\phi } + (i \sigma ^{12} + t_{3}) \cos \theta} \over
 { \sin \theta }}  \right )
$$

\noindent A characteristic feature of such a correlated choice of
frames in both these spaces is the explicit form of the total angular
momentum operator (the sum of orbital, spin, and isotopic ones)
$
J^{S.}_{1} = (l_{1} + {{(i \sigma ^{12} + t^{3}) \cos \phi } \over
{ \sin \theta }} ),
J^{S.}_{2} = (l_{2} + {{(i \sigma ^{12} + t^{3}) \sin \phi } \over
{ \sin \theta }} ),     J^{S.}_{3} = l_{3} ;
$
so that the present case entirely comes under the situation considered
by Pauli in [15]. The Pauli criterion allows here the following values
for $j: j = 0, 1, 2, 3,~\ldots $  In agreement  with a general procedure
 [8],  the
$\theta ,\phi $-dependence of composite multiplet wave function $\Psi _{jm}$
is to be built up from the Wigner $D$-functions:
 $D^{j}_{-m,\sigma } (\phi ,\theta ,0)$ ,
where the lower right index  $ \sigma $  takes the values from $(-1,0,+1)$,
which correlates with the explicit diagonal structure of the matrix
$ ( i\sigma ^{12} + t^{3} ) $ :
$$
\Psi _{\epsilon jm}(x) = {{ e^{-i \epsilon t}} \over r} \;
 [{ T_{+1/2} \otimes  F(r)} + { T_{-1/2} \otimes  G(r)} ]
\eqno(4)
$$

\noindent here (the fixed symbols $j$ and $-m$  in
${D}^{j}_{-m,\sigma } (\phi ,\theta ,0) $ are omitted)
$$
F(r) = \left( \begin{array}{l}
            f_{1}(r) D_{-1} \\
            f_{2}(r) D_{ 0} \\
            f_{3}(r) D_{-1} \\
            f_{4}(r) D_{ 0}
            \end{array}
       \right) ; \qquad
G(r) = \left( \begin{array}{l}
              g_{1}(r) D_{ 0} \\
              g_{2}(r) D_{+1} \\
              g_{3}(r) D_{ 0} \\
              g_{4}(r) D_{+1}
              \end{array}
       \right)
$$

\noindent further throughout the paper the temporal factor will be omitted.
Another essential feature of the given frame in  the
({\em Lorentz}) $\times$ ({\em isotopic})-space
is the appearance of the very simple expression for
the term that mixes up together two distinct   components  of  the
isotopic doublet (see eq. (3)). Moreover, it is evident at once,
that both these features are preserved, with no substantial
variations, when extending this particular problem to more
complex ones (with other given Lorentz and isotopic spins).

The separation of variables in the equation is accomplished
by a conventional $D$-function recursive relation fashion. Moreover,
only two relationships  from  the  enormous $D$~-function  technique
(see, for instance, in [23] ) are really needed in  doing  such  a
separation; they are
$$
{{\partial} \over {\partial \beta }} D^{j}_{mm'}(\alpha ,\beta ,\gamma ) =
+{1\over 2}  \sqrt {(j+m')(j-m'+1)} e^{-i\gamma } D^{j}_{m,m'-1}  -
$$
$$
 {1\over 2}  \sqrt {(j-m')(j+m'+1)} e^{+i\gamma } D^{j}_{m,m'+1} ;
$$
$$
{{m-m' \cos \theta} \over {\sin \theta}} D^{j}_{mm'}(\alpha ,\beta ,\gamma ) =
-{1\over 2}  \sqrt {(j+m')(j-m'+1)} e^{-i\gamma } D^{j}_{m,m'-1}  -
$$
$$
 {1\over 2}  \sqrt {(j-m')(j+m'+1)} e^{+i\gamma } D^{j}_{m,m'+1} \; .
$$

As known, an important case in theoretical investigation is
the electron-monopole system at the minimal value of the quantum
number $j$, so the case $j = 0$ should be considered in an
especially careful way.  In the chosen frame, it is the
independence on  $\theta ,\phi $-variables that sets the wave functions of
minimal $j$ apart from all other particle multiplet states (certainly,
functions $f_{1}(r)$ , $ f_{3}(r)$ , $ g_{2}(r) $ ,  $g_{4}(r)$
in the substitution (4) must be equated to zero at once). Besides,
the angular term  $\Sigma _{\theta ,\phi}$  in the wave equation is
eliminated effectively  due to the identity
$(i\sigma ^{12} + t^{3}) \Psi _{j=0} \equiv  0$.

The system of radial equations found by separation of
variables (it is 4 and 8 equations in the cases of $j = 0$ and $j > 0$
respectively) are rather complicated (they  cannot be considered
here in detail). They are simplified by searching a suitable
operator that could be diagonalized simultaneously with the
$ \vec j ^{2} , j_{3}$.
As well known, the usual space reflection ($P$-inversion) operator
for a bispinor field has to be followed by a certain discrete
transformation in the isotopic space, so that  a required quantity
could be constructed. Indeed, the solution of this problem, which
has been established up to date, is not general as much as
possible. For this reason, the question of reflection symmetry  in
the doublet-monopole system is reexamined here all again.
As a result we find that there are two different possibilities depending
on what type of external monopole potential is analyzed. So, in case of
the non-trivial potential, the composite reflection operator with required
properties is (apart from an arbitrary factor)
$$
\hat{N}^{S.} =  \hat{\pi } \otimes \hat{P}_{bisp.} \otimes  \hat{P} ,
\qquad   \hat{\pi } = + \sigma _{1}             \; .
\eqno(5a)
$$

\noindent A totally different situation occurs in case of the simplest
monopole potential. Now, a possible additional  operator suitable
for separating the variables depends on a certain arbitrary
complex parameter $A$ ($ e^{iA} \neq  0 $ ):
$$
\hat{N}^{S.}_{A} = \hat{\pi }_{A} \otimes \hat{P}_{bisp.} \otimes
\hat{P} \; , \qquad     \hat{\pi }_{A} = e^{iA \sigma _{3}} \sigma _{1}\; .
\eqno(5b)
$$

\noindent The same quantity  $A$ appears  also  in  expressions  for the wave
functions $ \Psi ^{A}_{\epsilon jm\delta}(t,r,\theta ,\phi)$
(the eigenvalues $ N_{A} = \delta  (-1)^{j+1} ; \delta  = \pm  1$) :
$$
\Psi ^{A}_{\epsilon jm\delta}(x) = [\; T_{+1/2} \otimes  F(x) \; + \;
                       \delta \; e^{iA} \; T_{-1/2} \otimes  G(x) \; ]
\eqno(5c)
$$

\noindent now the $g_{i}(r)$  from (4) are to be
$g_{1} = \delta  f_{4}$ , $g_{2} = \delta  f_{3}$ ,
$g_{3} = \delta  f_{2}$ , $g_{4} = \delta  f_{1}$ .

\subsection*{4. Case of special monopole field}

Now, the problem of simplest monopole field is  examined  more
closely.  The  system  of  radial  equations  specified  for  this
monopole potential is basically simpler than in general  case,  so
that the whole problem  including  the  radial  functions  can  be
carried to its complete conclusion:
$$
\Psi ^{A}_{\epsilon jm \delta \mu }(x) =  T_{+1/2}
\left(  \begin{array}{r}
            f_{1}(r) D_{-1} \\
            f_{2}(r) D_{ 0} \\
       \mu  f_{2}(r) D_{-1} \\
       \mu  f_{1}(r) D_{ 0}
\end{array}     \right)  \;\; + \;\;
  \mu \; \delta\;  e^{iA} \; T_{-1/2} \times \left(
           \begin{array}{r}
            f_{1}(r) D_{ 0} \\
            f_{2}(r) D_{+1} \\
       \mu  f_{2}(r) D_{ 0} \\
       \mu  f_{1}(r) D_{+1}
           \end{array}      \right)
\eqno(6)
$$

\noindent where $\delta  = \pm  1$ and $\mu  = \pm  1$, which are
independent of each  other; the quantum number $\mu$ relates to
the so-called {\em generalized} Dirac operator [17,18]
$\hat{K} = i \gamma^{0} \gamma^{3} \Sigma_{\theta \phi}$ ;
the  functions $f_{1}$  and $f_{2}$  satisfy a system of first order
differential equations. The $A$-ambiguity in the expansion (6)
indicates that this parameter $e^{iA}$  may be regarded as a quantity
measuring a violation of Abelicity in the composite non-Abelian
wave functions  $\Psi ^{A}_{\epsilon jm \delta \mu }(x)$. Significantly,
that the $e^{iA}$  must not be equal to zero; in other words,
the operator sets above
never lead to basis states without the $A$-violation. The two purely
Abelian multiplet states are  formally obtained from (5c) or (6) too:
it suffices to put $e^{iA} = 0$ or $\infty$, but these singular cases
are not covered by  operator sets under consideration. These two bound
values for $A$ provide us, in  a  sense, with the singular transition
points between the Abelian and non-Abelian theories.

On relating the expressions (6) with the Abelian  analogous solutions,
it follows that these non-Abelian functions are directly
associated with Abelian  monopole  ones  (those  were investigated by
many authors [1-8]):
$$
\Psi ^{A}_{\epsilon jm \delta \mu}(x) =
T_{+1/2} \otimes  \Phi ^{eg=-1/2}_{\epsilon jm\mu}(x) \;\; + \;\;
\mu\;  \delta \; e^{iA}\; T_{-1/2} \otimes  \Phi ^{eg=+1/2}_{\epsilon jm\mu }(x) \; ,
$$
$$
\Psi ^{A}_{\epsilon 0\delta }(x) =
T_{+1/2} \otimes  \Phi ^{eg=-1/2}_{\epsilon 0}(x)  \;\; +  \;\;
\delta \;  e^{iA} \; T_{-1/2} \otimes  \Phi ^{eg=+1/2}_{\epsilon 0}(x)  \; .
$$

\subsection*{5. On physical distinctions  between manifestations
of the Abelian and non-Abelian monopoles}

A key question of Section 5 concerns distinctions between the
Abelian and non-Abelian monopoles. We draw attention to that,  for
the distinguishable physical systems, namely, a free  isodoublet  with  no
external potentials and an isodoublet in the external monopole
field (whether a trivial or non-trivial one is meant), their spherical
symmetry operators $\vec{J}^{2}, J_{3}$   identically  coincide. In a
sequence, these isodoublet wave functions do not vary at all in
their dependence on angular variables $\theta , \phi $. This non-Abelian
wave function's property sharply contrasts with the Abelian one when
both electronic spherically symmetric wave functions and all the
symmetry operators undergo a significant transformation:
$$
j^{eg}_{1} =(l_{1}+{{(i\sigma ^{12} - eg) \cos \phi} \over{\sin \theta}}),
\qquad
j^{eg}_{2} =(l_{2}+{{(i\sigma ^{12} - eg) \sin \phi} \over{\sin \theta}}),
$$
$$
j^{eg}_{3} = l_{3}, \qquad
\Phi^{eg}_{\epsilon jm\mu} (t,r,\theta ,\phi) =
{e^{-i \epsilon t} \over r} =
\left( \begin{array}{r}
             f_{1} D^{j}_{eg-1/2}\\
             f_{2} D^{j}_{eg+1/2}\\
        \mu  f_{2} D^{j}_{eg-1/2}\\
        \mu  f_{1} D^{j}_{eg+1/2}\\
                            \end{array} \right)
$$
the value $eg = 0$  relates  to  the  free electronic  function.

We may note that one of the fundamental ideas underlying the theory
of the non-Abelian monopole is probably following: it being considered
as the external potential does not destroy the isotopic angular
structure of the particle multiplet wave functions. From this point of view,
it represents a certain analog of a spherically symmetric Abelian potential
$A_{\mu } = ( A_{0}(r), 0, 0, 0 )$
rather than an analog of the Abelian monopole potential
$A_{\mu } = (0, 0 , 0 , A_{\phi } = g \cos \theta)$.
Here we may draw attention to  that  the designation ``monopole''
in the non-Abelian terminology  anticipates an interpretation of
$W^{(a)}_{\mu }(x)$ as carrying, in a new situation, the essence of
the well-known Abelian monopole, although a real degree of their
similarity is going to be probably less than one should expect.

By the way, in both (Abelian and  non-Abelian) cases, monopole terms
involved in wave equations  vanish at  long  distances  (far  away
from $r = 0$), but in the non-Abelian problem there is no remaining
monopole manifestation through $\theta ,\phi $ - dependence, whereas
in the Abelian problem such an implicit monopole presence is still
conserved (as a result of $eg$~-displacement in Wigner $D$-functions
involved in wave functions). So we cannot get rid of the Abelian
monopole manifestation up to infinitely distance points, and  such
a property is removed far from what is familiar when a situation
is less singular (for instance, of electric charge or non-Abelian
monopole problem). In other words, the non-Abelian monopole, being
considered as external potential, provides a localized  object
(irrespective of whether the trivial or non-trivial monopole is
meant); this property is motivated only by its isotopic
structure. In contrast to this, the Abelian monopole is obviously
non-localizable object, and this finds its natural corollary in
giving rise the well-known difficulties on boundary conditions
(both  in classical and quantum mechanical scattering theory
[24-30]). One should emphasize that possible mutual series
expansions cannot be completely correct (in Abelian case) at
$\theta = 0$ and $\theta = \pi $,
so that the whole situation on the axis $x_3$  does not  conform  to
the basis superposition principle, whereas such a problem does not
arise at studying the analogous non-Abelian problem. The  intimate
explanation of this behavior involves the detailed examination  of
the boundary properties of corresponding  Wigner $D$-functions  and
cannot therefore be considered here in detail.

Thus, the whole multiplicity of the Abelian monopole manifestations seems
to be much more problematical than non-Abelian monopole's. Strictly
speaking, these two mathema\-ti\-cal situations are not related to each
other and an examination  for non-Abelian case will not lead up to solving
Abelian problems, but only being associated heuristically and thereby
thrown a further light  on each other. By way of illustration, we
consider the question of $P$-parity in both these theories
(see also in [31-37]).

\subsection*{6. $N$ -parity selection rules}

In the Abelian case (when $eg \neq 0)$, the monopole wave
functions cannot be proper functions of the usual space reflection
operator for the bispinor field. There exists only  the  following
relationship
$$
( \hat{P}_{bisp.} \otimes \hat {P} ) \; \Phi ^{eg}_{\epsilon jm\mu}(x)=
{e^{-i\epsilon t} \over r } \; \mu \;  (-1)^{j+1}
\left( \begin{array}{r}
          f_{1} D^{j}_{-m,-eg-1/2}\\
          f_{2} D^{j}_{-m,-eg+1/2}\\
          \mu f_{2} D^{j}_{-m,-eg-1/2}\\
          \mu f_{1} D^{j}_{-m,-eg+1/2}\\
                            \end{array} \right)
$$

\noindent
compare it with analogous one for the free wave functions:
$$
(\hat{P}_{bisp.} \otimes  \hat{P} ) \; \Phi ^{0}_{\epsilon jm\delta}(x) =
\delta \; (-1)^{j+1} \; \Phi ^{0}_{\epsilon jm\delta }(x)
$$

\noindent
A certain  diagonalized on these functions $\Phi ^{eg}_{\epsilon jm\mu }$
discrete operator obtains through multiplying the  usual $P$-inversion
bispinor operator by a formal one $\hat{\pi}$ which affects
the $eg$-parameter in the wave functions:
$\hat{\pi } \Phi ^{+eg}_{\epsilon jm\mu }(x)$ =$
\Phi ^{-eg}_{\epsilon jm\mu }(x)$. Thus, we have
$$
\hat{M} = \hat{\pi} \otimes \hat{P}_{bisp.} \otimes \hat{P} \; , \qquad
\hat{M} \; \Phi ^{eg}_{\epsilon jm \mu} (x) =
\mu\; (-1)^{j+1}\; \Phi ^{eg}_{\epsilon jm\mu}(x)
$$

\noindent but the latter fact does not allow us to obtain  $M$-parity
selection rules. Indeed, a matrix element for some physical observable
$\hat{G}^{0}(x)$  is  to  be
$$
\int \bar\Phi^{eg}_{\epsilon jm \mu}(x)
\hat{G}^{0}(x)    \Phi^{eg}_{\epsilon j'm'\mu'}(x) dV =
\int r^2 dr \int f(\vec{x})  d  \omega
$$

\noindent First we  examine the case $eg = 0,$ in order  to  compare  it
with the situation at $eg \neq  0$ . Let us relate
$f(-\vec{x})$ with $f(\vec{x})$.
Considering the equality  (and the same  with $j'm' \delta'$ )
$$
\Phi ^{0}_{\epsilon jm\delta }(-\vec{x} ) =
\hat{P}_{bisp.} \delta (-1)^{j+1} \Phi ^{0}_{\epsilon jm\delta }(\vec{x} )
\eqno(7a)
$$

\noindent we  get
$$
f(-\vec{x}) = \delta  \delta'(-1)^{j+j'+1}
\bar{\Phi}_{\epsilon jm\delta }(\vec{x})( \hat{P}^{+}_{bisp.}
\hat{G}^{0}(-\vec{x}) \hat{P}_{bisp.} ) \Phi _{\epsilon j'm'\delta'}(\vec{x})
$$

\noindent If $ \hat{G}^{0}(\vec{x})$  obeys  the equation
$$
\hat{P}^{+}_{bisp.} \hat{G}^{0}(-\vec{x}) \hat{P}_{bisp.} =
\omega ^{0} \hat{G}^{0}(\vec{x} )
\eqno(7b)
$$

\noindent here $\omega ^{0}$ defined to be $+1$ or $-1$ relates to
the scalar and pseudoscalar respectively, then the expression for
$f(-\vec{x}))$ above comes to
$f(-\vec{x}) = \omega  \delta  \delta' (-1)^{j+j'+1} f(\vec{x})$
that generates the well-known $P$-parity selection rules.

In contrast to everything just said, the situation at $eg \neq  0$
is completely different because any  equality in  the   form (7a)
does not exist there. So, there is no $M$-parity selection rules in
the presence of the Abelian monopole.  In  accordance  with  this,
for instance, an expectation value for the usual operator of space
coordinates $\vec{x}$  need not to equal zero and it follows this
(see in [36-37]).

Let us return to the non-Abelian problem when there exists  a
needed relationship
$$
\Psi_{\epsilon jm\delta} (- \vec{x}) =
( \sigma ^{2} \otimes \hat{P}_{bisp.})  \;
\delta \; (-1)^{j+1} \;  \Psi _{\epsilon jm\delta }(\vec{x})
\eqno(8a)
$$

\noindent owing to the $N$-reflection symmetry; so that
$$
f(-\vec{x}) = \delta\;  \delta'\; (-1)^{j+j'}
\bar{\Psi}_{\epsilon jm\delta} (\vec{x}) \otimes
[\; ( \sigma ^{2} \otimes \hat{P}^{+}_{bisp.})
\hat{G}(-\vec{x}) ( \sigma ^{2} \otimes \hat{P}_{bisp} )\; ] \;
\Psi _{\epsilon j'm' \delta'} (\vec{x})           \; .
$$

\noindent If a certain quantity $\hat{G}(\vec{x})$  which in  comparison
with a previous one depends on isotopic coordinates, obeys the condition
$$
( \sigma ^{2} \otimes  \hat{P}^{+}_{bisp.} )   \;
\hat{G} (-\vec{x})  \;
( \sigma ^{2} \otimes \hat{P}_{bisp.} ) = \omega \; \hat{G}(\vec{x})
\eqno(8b)
$$

\noindent  $\omega$ defined  to be $+1$ or $-1$, the  relationship above
converts  into
$
f(-\vec{x}) = \omega  \;  \delta \;  \delta' \; (-1)^{j+j'} f(\vec{x}) \; ,
$
that results in the evident $N$-parity selection rules. For
instance, applying this rule to $\hat{G}(\vec{x}) \equiv \vec{x}$ ,
we found
$
<\Psi _{\epsilon jm\delta} (x) \mid      \vec{x} \mid
 \Psi _{\epsilon jm\delta}(x) >   \sim
 [ 1 - \delta ^{2} (-1)^{2j} ] \equiv  0$ .
That vanishing may be also readily understood from  the  following
expansion of the matrix element
$$
< \Psi_{\epsilon jm\delta }(x) \mid \vec{x} \mid
  \Psi_{\epsilon jm\delta }(x) > =
  < \Psi^{-1/2}_{\epsilon jm }(x) \mid \vec{x} \mid
    \Phi^{-1/2}_{\epsilon jm }(x) > \; + \;
  < \Phi^{+1/2}_{\epsilon jm }(x) \mid  \vec{x} \mid
    \Phi^{+1/2}_{\epsilon jm }(x) >
$$

\noindent (where both isotopic components contribute equally  to  the  matrix
element)  and fitting  relationships
$
 <\Phi ^{\pm 1/2}_{\epsilon jm } (-\vec{x}) \mid - \vec{x} \mid
  \Phi ^{\pm 1/2}_{\epsilon jm }(-\vec{x}) > = -
 <\Phi ^{\mp 1/2}_{\epsilon jm } (\vec{x}) \mid  \vec{x} \mid
  \Phi ^{\mp 1/2}_{\epsilon jm }(\vec{x}) > \; .
$
From the preceding  it is evident that the solvable non-Abelian
problem of $N$-parity selection  rules does not guarantee that
another problem of $M$-parity (for the Abelian case) can
automatically be solved in analogous way.

One should notice that in the literature  there  are  several
suggestions how obtain a certain formal  covariance  of  the
monopole situation with respect to $P$-symmetry. Such attempts would
imply, for example, a pseudo  scalar  character  of  the  isolated
magnetic charge [31,33-35] or improvement [36] in understanding the
$P$-symmetry, which in turn provides  some genuine $P$-reflection
operator, etc. But, admittedly, all these solutions do not
permit to get over the non-existence of the discrete symmetry
selection rules for matrix elements at considering a single-particle
problem in  a fixed monopole potential.

\subsection*{7. New $N_{A}$-parities selection  rules}

First, by simple calculation, we detail explicit forms of
$\hat{N}_{A}$-operator in the unitary Dirac ($D.$) and Cartesian
($C.$)  gauges of isotopic space (see Suppl. A) :
$$
\hat{\pi}^{D.}_{A} = \left( \begin{array}{cc}
                         0       & -ie^{-iA} e^{-i\phi}  \\
                         +i e^{+iA} e^{+i\phi}  &     0
                     \end{array} \right) ; \qquad
\hat{\pi}^{C.}_{A} =
 (-i) exp \; [ iA\; \vec{\sigma}\; \vec{n}_{\theta,\phi}\; ]  \; .
\eqno(9)
$$

Now, we turn to the question how this complex characteristic
parameter $A$  can manifest itself. As a representative example, the above
problem of parity selection rule is investigated again, but now
depending on this  $A$-background.  For the composite physical
observable having inclusive constituent structure (its isotopic
content is separated out explicitly)
$$
G(\vec{x}) = \left( \begin{array}{cc}
         \hat{g}_{11}(\vec{x})  & \hat{g}_{12}(\vec{x}) \\
         \hat{g}_{21}(\vec{x})  & \hat{g}_{22}(\vec{x})
              \end{array} \right)
\times  G^{0}(\vec{x})
\eqno(10a)
$$

\noindent a natural definition of scalars and pseudoscalars relative to  the
$\hat{N}_{A}$-reflection occurs (compare it with (8b)):
$
( \hat{\pi }^{+}_{A} \otimes \hat{P}^{+}_{bisp.}) \;
\hat{G}(-\vec{x}) \;
( \hat{\pi }_{A} \otimes \hat{P}_{bisp.}) =
\Omega ^{A}\; \hat{G}(\vec{x})
$
or in more detailed form
$$
          \left( \begin{array}{cc}
e^{+i(A-A^{*})} \hat{g}_{22}(-\vec{x}) & e^{-i(A+A^{*})} \hat{g}_{21}(-\vec{x})\\
e^{+i(A+A^{*})} \hat{g}_{12}(-\vec{x}) & e^{-i(A-A^{*})} \hat{g}_{11}(-\vec{x})
          \end{array} \right)
[\; \hat{P}^{+}_{bisp.} \hat{G}^{0}(-\vec{x}) \hat{P}_{bisp.}\; ] =
\Omega^{A} \;  \hat{G}(\vec{x})
\eqno(10b)
$$

\noindent where $\Omega ^{A} = + 1$ or $-1$. For every given $A$ ,
the (10b) produces its own special limitations on composite scalars and
pseudoscalars, which are individualized by this  $A$. That is, the
different values of  $A$ lead to various concepts of scalars and
pseudoscalars respectively. Correspondingly, $N_{A}$-parity selection
rules arising in sequel for matrix elements (if an observable
belongs to either the $\Omega _{A} = + 1$  or $\Omega _{A}= - 1$
type)  differ basically from each other.

\subsection*{8. Parameter $A$ and isotopic ``chiral'' symmetry}

Section 8 is interested in the following question: Where does
the above A-ambiguity come  from?  Here,  we  can  note  that  all
different values for $A$  lead to the same whole functional  space;
each fixed  $A$  governs only the basis states:
$\Psi ^{A'}_{\epsilon jm\delta}(x) =
U(A',A) \Psi ^{A}_{\epsilon jm\delta }(x)$.
The explicit form of $U(A',A)$  in $S$.-gauge, is
$$
U_{S.}(A',A) = e^{+i(A'-A)/2}
     \left( \begin{array}{cc}
                          e^{-i(A'-A)/2}   &        0          \\
                          0                &   e^{+i(A'-A)/2}
     \end{array} \right)
\eqno(11a)
$$

\noindent correspondingly, in  $C$.-gauge, it is
$$
U_{C.}(A',A) = e^{+i(A'-A)/2}
\exp [\; (-i) \; {{A'-A}\over2} \;\vec{\sigma}\; \vec{n}_{\theta ,\phi}\; ]
\eqno(11b)
$$

\noindent In both cases (11a) and (11b) , the  second  factor  is the
2-spinor transformation lying in the  3-dimensional  complex  rotation
group $SO(3.C)$. In addition, as readily verified, the operations
$\hat{N}$ and $\hat{N}_{A}$ turn out to be connected by the relation
$$
\hat{N}^{S.}_{A} = U_{S.}(A,0) \hat{N}^{S.} U^{-1}_{S.}(A,0)\;  ;\qquad
\hat{N}^{C.}_{A} =
U_{C.}(A,0) \hat{N}^{C.} U^{-1}_{C.}(A,0)  \; .
\eqno(12)
$$

\noindent  In this connection, the question  arisen  is  how  to
evaluate contrasting these two discrete operators
$\hat{N}^{S.}$  and $\hat{N}^{S.}_{A}$ .
 So, we have to give more attention to this relationship
$\hat{N}^{S.}_{A}$  and $\hat{N}^{S.}$.
To clearing up this matter, as turned out, it suffices to draw consistently
distinction between two situations. The first one  concerns the sets
$[ \hat{J}_{i}, \hat{N}^{S.}]$  and  $[\hat{J}_{i}, \hat{N}^{S.}_{A}] $
when they are considered as different but equivalent realizations of the
same given representation of the group $SO(3.R)$. The second
relates to case when these two operator sets are regarded as the
physical observables at the same  physical  system  of fixed
Hamiltonian:
$
[\hat{J}_{i}, \hat{N}^{S.}]^{H}\; ,   \;
[\hat{J}_{i}, \hat{N}^{S.}_{A}]^{H}\; ;
$
and then they are physically distinguishable as generating further
different  basis wave functions. An analogy with a more familiar example
of the  Dirac  massless field can be called  [18], when the
{\em complex}  chiral  symmetry transformation
$$
\left( \begin{array}{c}
                     \xi  '(x)  \\
                     \eta '(x)
                                  \end{array} \right) =
\left( \begin{array}{cc}
                      1     &   0   \\
                      0     &   Z
                                   \end{array} \right)
\left( \begin{array}{c}
                     \xi  (x)  \\
                     \eta (x)
                                  \end{array} \right) ; \qquad
\Psi' (x) = e^{iA/2} \exp (i{A\over2} \gamma^{5}) \Psi(x)
$$

\noindent ($Z = e^{iA}$)  leads  us  to  use  (alternatively) the generalized
$P$-reflection bispinor  operator:
$
e^{iA \gamma ^5} \; \hat{P}_{bisp.} \otimes \hat{P} \; .
$
Moreover, its explicit form remains basically unchanged at  translating  the
spheric tetrad basis  into  Cartesian's  ($\hat{P}^{C.}_{bisp} = i \gamma^{0}$)
as the $\gamma ^5$ and the gauge  transformation  involved  are commutative
with each other. Such a condition is not realized for the non-Abelian
monopole-electron system, so the Schwinger  expression of the
$\hat{N}^{S.}_{A}$, after transformation to the Cartesian isotopic basis,
takes the form in which  the $\theta, \phi$-dependence appears explicitly:
$$
\hat{N}^{S.}_{A} = ( e^{iA \sigma_3}  \; \hat{\pi } \otimes
                 \hat{P}_{bisp.}) \otimes  \hat{P} \; ; \qquad
\hat{N}^{C.}_{A} =
[\;(-i) \;\exp( i A \;\vec {\sigma}\; \vec{n}_{\theta,\phi}\; ) \otimes
\hat{P}_{bisp.} \otimes \hat{P} ]               \; .
\eqno(13)
$$

As an additional remark, it should be mentioned that starting solely
from the usual Cartesian isotopic formulation, we could not have found out
(with great probability) such an $A$-ambiguity as in (13) and also we could
not have disclosed the existence itself of a  possible isotopic
chiral symmetry in non-Abelian nonopole-electron system. Thus, an~incidental
choice of a~basis (both in the~isotopic  and  bispinor spaces) results in
unexpected possibilities in the characterization of this system.

\subsection*{9.Complex values of the $A$ and a~collision between the~quantum
    mechanical superposition principle and self-conjugacy requirement}

In this Sec. 9, let us look closely at  some  qualitative peculiarities
of the above considered  $A$-freedom  placing  special notice to the~division
of $A$-s into the real and complex ones. It is convenient to work at
this matter in the Schwinger unitary basis. Recall that the $A$-freedom
tell us  that  simultaneously  with
$\hat{H}, \vec{j}^{2} , \hat{j}_{3}$,
else one discrete operator $ \hat{N}_{A} $  that  depends  generally
on a complex number $A$  can be diagonalized on the wave  functions.
Correspondingly, the basis functions associated with the complete set
 ( $ \hat{H} ,\vec{j}^{2}, \hat{j}_{3},\hat{N}_{A}$ )
besides being certain determined functions of the relevant quantum
numbers $(\epsilon , j, m, \delta  )$ , are subject to the $A$ :
$$
\Psi ^{A}_{\epsilon jm\delta }(x) =
              [ \; T_{+1/2} \otimes  \Phi ^{+}_{\epsilon jm}(x) \; + \;
\delta e^{iA} \;   T_{-1/2} \otimes  \Phi ^{-}_{\epsilon jm}(x) \; ]  \; .
\eqno(14a)
$$

\noindent In other words, all different values of this  $A$ lead to different
quantum-mechanical bases of the system. There exists a~set of
possibilities, but one can relate every two of them  by means of
a~respective linear transformation. For example, the states
$\Psi ^{A}_{\epsilon jm\delta }(x)$
decompose into the following linear combinations of the initial  states
$\Psi ^{A=0}_{\epsilon jm\delta }(x)$
(further,  this  $A=0$ index  will  be omitted):
$$
\Psi ^{A}_{\epsilon jm\delta }(x) =
\left [\; {{1 + \delta  e^{iA} } \over 2}\; \Psi _{\epsilon jm,+1} \;+ \;
  {{1 - \delta  e^{iA} } \over 2} \; \Psi _{\epsilon jm,-1} \right ] \; .
\eqno(14b)
$$

\noindent One should give heed to that, no matter  what an $A$   is
(either real or complex one), the new states (14b) being linear
combinations of the initial states are permissible as well as
old ones. This added aspect of the~allowance of the complex values
for $A$ conforms to the quantum-mechanical superposition principle,
the latter presupposes that arbitrary complex coefficients $c_{i}$
in a~linear combination of some basis states
$\Sigma  c_{i} \Psi _{i}$     are acceptable.

However, an essential and subtle distinction between real and
complex $A$-s comes straightforward to light  as we turn to the
matter of normalization and orthogonality for
$\Psi ^{A}_{\epsilon jm\delta }(x)$. An elementary calculation gives
$$
< \Psi^{A}_{\epsilon jm,\delta} \mid \Psi^{A}_{\epsilon jm,\delta}  >
= { {1 + e^{i(A-A^{*}) }} \over 2}  \; ;  \;\;
< \Psi^{A}_{\epsilon jm,\delta} \mid \Psi^{A}_{\epsilon jm,-\delta} >
= { {1 - e^{i(A-A^{*}) }} \over 2}
\eqno(15)
$$

\noindent i.e. if $A \neq  A^{*}$  then the normalizing condition
for   $\Psi ^{A}_{\epsilon jm\delta }(x)$    does
not coincide with that for   $\Psi _{\epsilon jm\delta }(x)$,
and what is more, the states
$\Psi ^{A}_{\epsilon jm,-1}(x)$  and  $\Psi ^{A}_{\epsilon jm,+1}(x)$
are not mutually orthogonal. The latter means that we face here
the~non-orthogonal basis in Hilbert space and the pure imaginary part of
the~$A$ plays a~crucial role in the~description of its non-orthogonality
property.

The {\em oblique} character of the  basis $\Psi ^{A}_{\epsilon jm\delta }(x) $
 (if $A \neq A^{*}$ )
exhibits its very essential qualitative distinction from
{\em perpendicular} one for $\Psi _{\epsilon jm\delta }(x)$. But, in a~sense,
the existence of the~non-orthogonal bases in the~Hilbert space represents
a~direct consequence of the~quantum-mechanical superposition principle. By
the way, for this reason, a~prohibition against complex $A$-s  could
be  partly a~prohibition against the conventional superposition
principle too; since all complex values for $A$, having forbidden,
imply specific limitations on two coefficients in (14b); but those are not
presupposed by the~superposition principle itself.

Up to this point, the~complex $A$-s seem to be good as well as
the~real ones. Now, it is the~moment  to  point  to some clouds
handing over this part of the subject. Indeed, as readily
verified, the operator $\hat{N}_{A}$ does not represent a~self-conjugated
(self-adjoint) one\footnote{The author is grateful to Dr. E.A.Tolkachev
for pointing out that it is so}
$
< \hat{N}_{A} \Phi (x) \mid  \Psi (x) > =
< \Phi (x) \mid  e^{i(A-A^{*}) \sigma_{3}} \; \hat{N}_{A} \Psi (x) >\; .
$
It is understandable that this (nonself-conjugacy) property
correlates with the~above-mentioned nonorthogonality
conditions (as well known, a~self-conjugated operator entails
both real its eigenvalues and the orthogonality of its
eigenfunctions). As already noted, the  eigenvalues of $\hat{N}_{A}$  are
real ones and this conforms to the~general statement that
all inversion-like operators possess the property of the~kind: if
$\hat{G}^{2} = I$  then $\lambda $  is a real number , as
 $\hat{G} \Phi _{\lambda } = \lambda  \Phi _{\lambda } )$.

So, we have got into a clash: whether one has  to  reject  all
complex values  for   $A$   and  thereby  violate  the  one  quantum
mechanical principle of major generality (of superposition) or whether it
is remain to accept all complex $A$-s as well as real ones and
thereby , in turn, stretch another  quantum-mechanical  regulation
about  the~self-conjugate character of physical quantities.
Thus, the physical system  under consideration exhibits inself
a~logical collision between two conventional quantum
propositions of principal significance. In the author's opinion,
one should accord the primacy of the general superposition
principle over the self-adjointness requirement. In  support of this
point of view , there exist some physical grounds.
Indeed, recall the quantum-mechanical status of all inversion-like quantities:
they serve always to distinguish two quantum-mechanical states. Moreover,
to those  quantum  variables there not correspond any classical variables;
the latter correlates with that any classical apparatus  measuring those
discrete variables does not exist at all. In contrast to this, one
should recollect  why  the~self-adjointness  requirement  itself  was
imposed on physical quantum operators. The reason is that such operators
imply all their eigenvalues to be real. Besides, that limitation
on physical quantum  variables had been put, in the first place,
for quantum variables having their classical counterparts (with
the continuum of classical  values measured). And after this,  in
the second place, the  discrete quantities  such  as $P$-inversion
and like it were tacitly incorporated into a~set of self-adjoint
mathematical operations, as a {\em natural} extrapolation.
But one should notice (and the~author inclines to place a~special emphasis
on this) the fact that the~single relation  $\hat{N}^{2}_{A} = I$
is completely sufficient that the~eigenvalues of $\hat{N}_{A}$ to be real. In
the light of this, the~above-mentioned automatic incorporation  of
those discrete operators into a~set of self-adjoint ones does not
seem inevitable.
But admitting this, there is  a~problem to
solve: what is the meaning of complex expectation  value  of  such
non self-adjoint discrete operators; since,  evidently,  the
conventional formula
 $<\Psi  \mid  \hat{N}_{A} \mid  \Psi  >$
provides  us  with  a~complex value. Indeed, let
$\Psi (x)$  be  $\Psi (x) = [ m \Psi _{+1}(x) + n \Psi _{-1}(x)]$, then
$$
< \Psi \mid \hat{N}_{A} \mid \Psi > =
< m \; \Psi_{+1}(x) \; + \;  n\; \Psi_{-1}(x) \mid m \; \Psi_{+1}(x)\;  - \;
n\; \Psi _{-1}(x) > =
\eqno(16)
$$
$$
\left   [\;  ( m^{*} m - n^{*} n )\;  {{1+e^{i(A-A^{*})}} \over 2} \; +\;
  ( n^{*} m - n m^{*} )\;  {{1-e^{i(A-A^{*})}} \over 2}\; \right  ]    \; .
$$

\noindent Must one be skeptical about those complex   $ \bar{N}_{A}$ ,
or treat them as physically acceptable quantities? Let us examine this
problem in more detail. It is reasonable to begin with an~elementary
consideration of the~measuring procedure of
the~$\hat{N} = \hat{N}_{A=0}$. Let  a~wave function $\Psi (x)$ decompose
into the combination
$$
\Psi (x) = [\; e^{i\alpha } \; \cos^{2}  \Gamma  \; \Psi _{+1}(x) \; + \;
             e^{i\beta  } \; \sin^{2} \Gamma \; \Psi _{-1}(x)\; ]
\eqno(17a)
$$

\noindent where $\alpha $  and $\beta \in  [ 0 , 2 \pi ]$,
and $\Gamma  \in  [ 0 , \pi /2 ]$. For  the $\hat{N}$
expectation value, one gets
$$
\bar{N} = < \Psi  \mid  \hat{N} \mid  \Psi  > =
(-1)^{j+1}\; ( \; \cos^{2} \Gamma  \; - \; \sin ^{2} \Gamma \; ) =
(-1)^{j+1} \; \cos 2\Gamma   \; .
\eqno(17b)
$$

\noindent From (17b) , one can  conclude  that  the $\bar{N}$ having  measured,
provides us only with the information about the parameter $\Gamma $  at (17a) ,
but does not furnish any information on the  phase  factors $e^{i\alpha }$
and $e^{i\beta }$ (or  their  relative  factor $e^{i(\alpha -\beta )}$).
  Such   an
interpretation of measured  $\bar{N}$  as receptacle of the quite definite
information about superposition coefficients in the  decomposition
(17a) represents one and only physical meaning of  the $ \bar{N}$.

Now, returning to the case of $\hat{N}_{A}$ operation, one  should  put
an analogous question  concerning the   $\bar{N}_{A}$ .
The required question is: what information about $\Psi (x)$  can be
extracted from the~measured   $\bar{N}_{A}$. It is convenient to rewrite
the~above function  $\Psi (x)$ as a~linear
combination of functions $\Psi ^{A}_{\epsilon jm,+1}$ and
$\Psi ^{A}_{\epsilon jm,-1}$.  Thus inverting the relations (14b) we get
$$
\Psi _{\epsilon jm,+1} =
\left [\; {{1 + e^{-iA}} \over 2} \; \Psi ^{A}_{\epsilon jm,+1} \;+ \;
 {{1 - e^{-iA}} \over 2} \; \Psi ^{A}_{\epsilon jm,-1} \; \right ]\;  ,
$$
$$
\Psi _{\epsilon jm,-1} =
\left [ \; {{1 - e^{-iA}} \over 2}\;  \Psi ^{A}_{\epsilon jm,+1} \; + \;
 {{1 + e^{-iA}} \over 2} \; \Psi ^{A}_{\epsilon jm,-1}\; \right ]
$$

\noindent and then $\Psi (x)$ takes the form  (the  quantum  numbers
$\epsilon ,j,m$   as fixed ones are omitted)
$$
\Psi (x) = \left [\; (\; e^{i\alpha} \; \cos \Gamma\;  {{1 + e^{-iA}} \over 2} \;+\;
             e^{i\beta } \;\sin \Gamma \; {{1 - e^{-iA}} \over 2} \;)
\; \Psi ^{A}_{+1}(x)  +    \right.
\eqno(18a)
$$
$$
\left.
(\; e^{i\alpha} \; \cos \Gamma \;  {{1 - e^{-iA}} \over 2} \; + \;
             e^{i\beta } \; \sin \Gamma \;   {{1 + e^{-iA}} \over 2} \;)
\; \Psi ^{A}_{+1}(x) \; \right  ] =
  [\;  m \; \Psi ^{A}_{+1}(x) \; +\;  n\; \Psi ^{A}_{-1}(x)\; ] \; .
$$

\noindent Although  the quantity  $A$   enters the expansion (18a),
but really
$\Psi (x)$ only contains three arbitrary parameters: those are
$\Gamma  , e^{i\alpha }$,
and $e^{i\beta }$. After simple calculation one gets
$$
\bar{N}_{A} = < \Psi  \mid  \hat{N}_{A} \mid  \Psi  > = (-1)^{j+1}  \;
(\; \rho\; \cosh g \; + \; i \sigma \; \sinh g \;  ) \; ,
\eqno(18b)
$$
$$
\rho = \cos 2 \Gamma\;  \cos f \; + \; \sin 2\Gamma \; \sin f \; \sin (\alpha -\beta )\; ,
$$
$$
\sigma = - \cos 2 \; \Gamma \sin f \;+\; \sin 2\Gamma \;\cos f\; \sin (\alpha -\beta )
$$

\noindent where  $f$  and $g$  are real  parameters defined by $A = f + i g$ .
Examining this expression, one may single out four particular
cases for separate consideration. Those are :
$$
1. \qquad g = 0 , f = 0 \;: \qquad
\bar{N}_{A} = (-1)^{j+1} \cos 2\Gamma
\eqno(19a)
$$

\noindent here, the  $\bar{N}$ only fixes $\Gamma $, but $e^{i(\alpha -\beta )}$
remains indefinite.
$$
2.   g = 0 , f \neq  0\;: \qquad  \bar{N}_{A} =
(-1)^{j+1} [ \; \cos 2\Gamma \cos f + \sin 2\Gamma \sin f \sin (\alpha -\beta ) \; ]
\eqno(19b)
$$

\noindent here, the   $\bar{N}_{A}$  measured does not fixes
$\Gamma $ and $(\alpha - \beta )$, but only imposes a~certain limitation  on
both these parameters.
$$
3.  \; g \neq  0 \; , \;  f = 0 \; : \qquad   \bar{N}_{A} =
(-1)^{j+1}\; [\; \cos 2 \Gamma \; \cosh g \; +  \;
 i \sin 2 \Gamma \; \sin (\alpha -\beta) \; \sinh g \; ]
\eqno(19c)
$$

\noindent here, the  $\bar{N}_{A}$  determines both $\Gamma $  and
$(\alpha - \beta )$;  and  thereby
this complex  $\bar{N}_{A}$  is a physical quantity being quite interpreted.
Finaly, for the fourth case
$$
4.\; g \neq  0 , f \neq  0  \; :
\cos 2\Gamma = (\; \rho \; \cos f \;-\; \sigma \sin f \; )\; ,
$$
$$
\sin 2\Gamma \; \sin (\alpha -\beta ) = ( \; \rho \; \cos f\; +\; \sigma \sin f \;)
\eqno(19d)
$$

\noindent i.e. the complex $\bar{N}_{A}$  also gives some information
about $\Gamma $   and $(\alpha - \beta )$ and therefore has character of
a  physically  interpreted  quantity.

\end{document}